# Enhanced cosmic-ray flux toward ζ Persei inferred from laboratory study of $H_3^+$–$e^-$ recombination rate

B. J. McCall*[+], A. J. Huneycutt*, R. J. Saykally*, T. R. Geballe[±], N. Djuric[||], G. H. Dunn[||], J. Semaniak[¤], O. Novotny[¤§], A. Al-Khalili[¶], A. Ehlerding[¶], F. Hellberg[¶], S. Kalhori[¶], A. Neau[¶], R. Thomas[¶], F. Österdahl[‡], and M. Larsson[¶]

*Department of Chemistry, University of California at Berkeley, Berkeley, California 94720, USA

[+]Department of Astronomy, University of California at Berkeley, Berkeley, California 94720, USA

[±]Gemini Observatory, 670 North A'ohoku Place, Hilo, Hawaii 96720, USA

[||]JILA, University of Colorado and National Institute of Standards and Technology, Boulder, Colorado 80309, USA

[¤]Institute of Physics, Świetokrzyska Academy, 25 406 Kielce, Poland

[§]Department of Electronics and Vacuum Physics, Faculty of Mathematics and Physics, Charles University Prague V Holesovickach 2, Prague 8, Czech Republic

[¶]Department of Physics, SCFAB, Stockholm University, S-106 91 Stockholm, Sweden

[‡]Manne Siegbahn Laboratory, Stockholm University, S-104 05 Stockholm, Sweden

The $H_3^+$ molecular ion plays a fundamental role in interstellar chemistry, as it initiates a network of chemical reactions that produce many interstellar molecules.[1,2] In dense clouds, the $H_3^+$ abundance is understood using a simple chemical model, from which observations of $H_3^+$ yield valuable estimates of cloud path length, density, and temperature.[3,4] On the other hand, observations of diffuse clouds have suggested that $H_3^+$ is considerably more abundant than expected from the chemical models.[5-7] However, diffuse cloud models have been hampered by the uncertain values of three key parameters: the rate of $H_3^+$ destruction by electrons, the electron fraction, and the cosmic-ray ionisation rate. Here we report a direct experimental measurement of the $H_3^+$ destruction rate under nearly interstellar conditions. We also report the observation of $H_3^+$ in a diffuse cloud (towards ζ Persei) where the electron fraction is already known. Taken together, these results allow us to derive the value of the third uncertain model parameter: we find that the cosmic-ray ionisation rate in this sightline is forty times faster than previously assumed. If such a high cosmic-ray flux is indeed ubiquitous in diffuse clouds, the discrepancy between chemical models and the previous observations[5-7] of $H_3^+$ can be resolved.

The dissociative recombination (DR) of $H_3^+$, i.e. the exothermic reaction $H_3^+ + e \rightarrow H + H + H$ or $H_2 + H$, has been one of the most controversial topics in the field of ion physics,[8,9] as different experimental methods have disagreed by a factor of 10,000 on the value of the thermal rate coefficient. In addition, it was only in the past few years that a plausible theoretical mechanism (the "indirect process"[10]) was used to explain the DR of $H_3^+$ in its lowest vibrational level[11]. However, those calculations gave a cross section[12] 1,000 times smaller than those measured by the ion storage rings CRYRING[13] (Sweden), ASTRID[14] (Denmark), and TARN II[15] (Japan), which all gave essentially the same result. Calculations[16] accounting for the symmetry breaking of the equilateral triangle $H_3^+$ ion by the incoming electron were in better accord with the experiments, and a more detailed theoretical treatment[17] has now provided a major step forward in terms of agreement, although important discrepancies with experiment remain to be resolved.

Meanwhile, a series of experiments at the Test Storage Ring in Germany[18,19] and at CRYRING[20] suggested that, while $H_3^+$ undergoes complete vibrational relaxation by spontaneous emission of radiation in storage ring experiments, the high rotational temperature of $H_3^+$ affected the measured values of the DR rate coefficient in the earlier experiments.[13-15] These experiments thus made clear that a measurement of a thermal rate coefficient applicable to interstellar clouds would have to be performed on rotationally cold $H_3^+$. This represented an unparalleled challenge to the ion storage ring technique, which has traditionally used hot filament or hollow cathode ion sources that produce ions at much higher rotational temperatures than those of diffuse clouds.

The challenge of producing rotationally cold $H_3^+$ ions has been addressed by the development of a new supersonic expansion ion source (see Fig. 1a). The measured rate coefficient for the DR of $H_3^+$ using this new source is shown in Fig. 2. The general shape is similar to that reported in earlier studies,[13,14] but these new data show a number of resonances between 1 meV and 1 eV. While hints of resonance structures have been reported previously,[20] the present experiment is the first to use rotationally cold ions, which apparently enhance the structure. Figure 3 shows the derived rate coefficient for a thermal electron distribution ($k_e$), which is about 40% lower than that measured in storage ring studies of rotationally hot $H_3^+$ ions[13,14].

While different ion storage ring experiments have consistently yielded about the same value of $k_e \sim 1 \times 10^{-7}$ cm$^3$s$^{-1}$ at 300 K (aside from the present experiment, which for the first time uses rotationally cold ions), afterglow experiments, which indirectly infer $k_e$ from the removal of electrons in a decaying plasma of hydrogen and noble gases, are more difficult to interpret. Different afterglow measurements[23] since 1984 have yielded values of $k_e$ (at 300 K) which range from $<10^{-11}$ to $\sim 2 \times 10^{-7}$ cm$^3$s$^{-1}$. The reason for this wide variance in afterglow measurements is still not understood, although it may be related to the complicated modelling and analysis that goes into extracting the value of $k_e$ from the experimental measurements. In contrast, the storage ring experiments are conceptually very simple and require a comparatively straightforward analysis to determine $k_e$, so we consider the present experiment to provide a much more robust measurement of the appropriate value of $k_e$ for interstellar conditions. It is important to note two differences between the present experiment and interstellar conditions: the presence of high electron number densities [$n(e^-) \sim 10^7$ cm$^{-3}$] and magnetic fields [$\sim 300$ Gauss]. Because these are inherent to the storage-ring technique and cannot be avoided, it is imperative that our results be confirmed by theoretical calculations. While there are still some unexplained differences between theory and our experiment, the most recent calculations[17] yield a value of $k_e$ ($\sim 2 \times 10^{-7}$ cm$^3$s$^{-1}$ at 40 K) that is similar to our results. Although we cannot explain the wide range in afterglow results, the preponderance of the evidence suggests that we have determined the appropriate value of $k_e$ for interstellar conditions, and this fact compels us to examine the implications of adopting this new value for modelling $H_3^+$ chemistry in diffuse clouds.

The chemical model for $H_3^+$ in diffuse clouds is quite simple, in contrast to the involved models needed for most other molecules. $H_3^+$ is formed when $H_2$ is ionised by a cosmic ray (a high energy proton or electron) to form $H_2^+$, which then undergoes a very rapid ion-neutral reaction with another $H_2$ to form $H_3^+$ and an H atom. The rate of formation can be expressed as $\zeta n(H_2)$, where $\zeta$ is the cosmic-ray ionisation rate and $n(H_2)$ is the number density of $H_2$. The destruction of $H_3^+$ in diffuse clouds is dominated by DR, and the rate of this process can be expressed as $k_e n(e^-) n(H_3^+)$. In steady state[7], the $H_3^+$ number density can then be written as $n(H_3^+)=(\zeta/k_e)n(H_2)/n(e^-)$.

Astronomical absorption spectra, however, yield direct information only about the absorber's column density (the integral of the number density along the line of sight), not the number density itself. We assume that the ratio of $n(H_2)/n(e^-)$ is approximately constant throughout the cloud, and replace it by the ratio of column densities $N(H_2)/N(e^-)$. Given this assumption, $n(H_3^+)$ is itself a constant, and we write $N(H_3^+)=n(H_3^+) L$, where L is the absorption path length. Substitution into the above equation then yields the relation $\zeta L=k_e (N(e^-)/N(H_2)) N(H_3^+)$.

Here we report the observation of $H_3^+$ in the diffuse cloud towards ζ Persei (Fig. 1b), which represents a major breakthrough, as this is the first detection of $H_3^+$ in a sightline where the electron fraction can be estimated. Assuming that most electrons come from the photo-ionisation of carbon, we can use existing ultraviolet measurements of molecular hydrogen[24] and ionised carbon[25] to derive the electron fraction $N(e^-)/N(H_2) = 3.8 \times 10^{-4}$. Combining this electron fraction, our observed $H_3^+$ column density, and the value $k_e = 2.6 \times 10^{-7}$ cm$^3$s$^{-1}$ from the CRYRING experiment (Fig. 3), we find ζL≈8,000 cm s$^{-1}$ for this sightline.

There is little direct observational information about the value of either ζ or L. It has generally been assumed that the value of ζ in diffuse clouds is the same as in dense clouds, where various measurements suggest a value of $ζ = 3 \times 10^{-17}$ s$^{-1}$. The absorption path length can be indirectly estimated by dividing the total column density of hydrogen nuclei by an assumed number density. The total column density towards ζ Persei is $N_H \equiv N(H) + 2N(H_2) \sim 1.6 \times 10^{21}$ cm$^{-2}$ (adopting $N(H) = 6.3 \times 10^{20}$ cm$^{-2}$, Ref. 26). The number density can be estimated from the rotational excitation of the CO and $C_2$ molecules, as well as from more detailed cloud models that consider the $H/H_2$ ratio[27]. While each of these methods has its uncertainties, they are all consistent with an average density of $<n_H> = 150–500$ cm$^{-3}$. If we adopt the density of $<n_H> = 250$ cm$^{-3}$, we obtain a path length of L=2.1 pc.

Together, the "canonical" value of ζ and the best estimate of L yield ζL=200 cm s$^{-1}$. However, the measured $H_3^+$ column density implies ζL=8,000 cm s$^{-1}$. If we adopt the canonical value of ζ, we would have to accept a very long path length of L=85 pc and a low density of $<n_H> = 6$ cm$^{-3}$. On the other hand, if we adopt the canonical path length, we must increase the ionisation rate to $ζ = 1.2 \times 10^{-15}$ s$^{-1}$. Given the evidence pointing to a short path length, and the fact that ζ is essentially unconstrained in diffuse clouds, we consider the higher ionisation rate to be the more likely explanation. A higher value of ζ for diffuse clouds may also find some support from considerations of the abundance of the OH molecule[28]. A more detailed discussion of the ζ Persei sightline, as well as other observations of $H_3^+$ in diffuse clouds, will be presented in a forthcoming paper.

The high value of ζ suggested by the CRYRING measurements and the ζ Persei observations can be reconciled with the lower ionisation rate in dense clouds by postulating the existence of a large flux of low energy cosmic rays that can penetrate diffuse but not dense clouds. Such a cosmic ray flux would bring the models and observations of $H_3^+$ into agreement, but would also have far-reaching implications for the chemistry and physics of interstellar clouds. From a chemical perspective, a higher value of ζ would proportionally increase the number densities of oxygen compounds such as OH, as well as affecting the relative abundances of deuterium-bearing molecules. The high number density of $H_3^+$ in diffuse clouds would also increase the rate of proton-transfer reactions, which play the key role in the formation of complex molecules in dense clouds but have been considered relatively unimportant in diffuse clouds. From a physical perspective, a higher cosmic ray ionisation rate represents an additional heating source for interstellar gas, and could have a significant impact on the thermal balance of the warm neutral medium[29]. Clearly, further observations of $H_3^+$, $H_2$, and $C^+$ in other diffuse cloud sightlines are needed to determine if the ζ Persei sightline is unusual, or whether a large low energy cosmic ray flux indeed pervades the galaxy.

The idea of using a supersonic expansion source for DR measurements originated during conversations between BJM and C. M. Lindsay, and the experiment was designed by BJM, AJH, and ML. The supersonic expansion ion source was built and spectroscopically characterized by AJH and BJM in the laboratory of RJS in Berkeley. The DR measurements in Stockholm were carried out by all of the authors except RJS and TRG. The UKIRT observations were obtained by BJM and TRG. BJM and AJH wish to thank K. Wilson and C.-Y. Chung for their assistance in the laboratory, D. Lucas for the loan of equipment, and H. Chan and E. Granlund for their superb support. BJM wishes to thank T. Oka, L. M. Hobbs, D. G. York, and T. P. Snow for helpful conversations about the ζ Persei sightline. We thank C. H. Greene for providing us with the results of his calculations in advance of publication. ML thanks D. Zajfman for information on TSR results prior to publication.

Correspondence and requests for materials should be addressed to BJM (e-mail: bjmccall@astro.berkeley.edu).

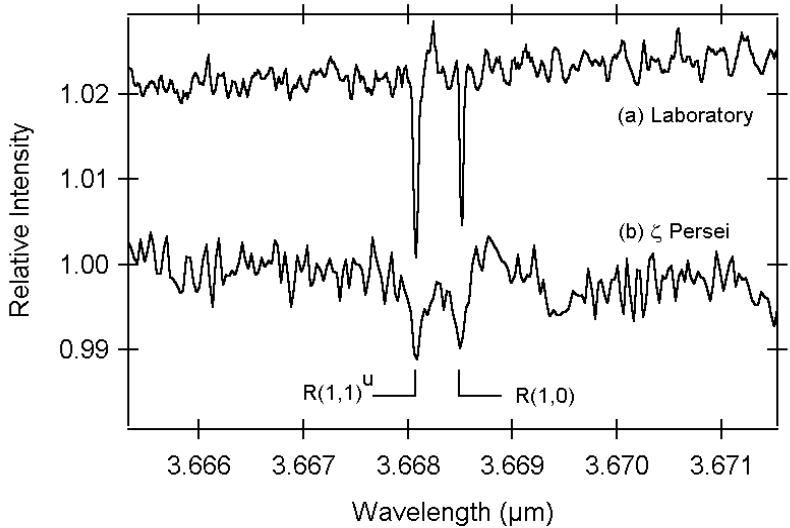

Figure 1: Spectra of two $H_3^+$ transitions arising from the two lowest rotational levels, which are the only levels with significant population in diffuse clouds. $R(1,1)^u$ originates from the lowest *para* level (J=1, K=1), while R(1,0) comes from the lowest *ortho* level (J=1, K=0). Note that the (J=K=0) level is forbidden by the Pauli principle. For energy level diagrams and a description of the $H_3^+$ notation, see ref. 21. (a) Cavity ringdown[22] laser absorption spectrum of the supersonic expansion ion source. In this source, $H_3^+$ number densities of ca. $10^{11}$ cm$^{-3}$ were produced in a direct current discharge plasma downstream of a 500 μm pinhole through which hydrogen gas (at 2.5 atmospheres) expanded supersonically into a vacuum. Depending on conditions, the rotational temperature of the plasma was 20–60 K, based on the relative intensities of the two transitions. This laboratory spectrum has been smoothed and multiplied by a factor of 1000 for clarity. (b) Spectrum of diffuse cloud towards ζ Persei obtained using the CGS4 infrared spectrometer at the United Kingdom Infrared Telescope (UKIRT) on the night of 2001 September 5 UT using our standard observing techniques[7]. The total on-source integration time was 40 minutes, and the ζ Persei spectrum has been divided by that of a standard star (BS 936) to remove absorption lines from the Earth's atmosphere. The spectrum is displayed in rest wavelengths, and indicates an $H_3^+$ column density (the integral of the number density along the line of sight) of $N(H_3^+)=8\times10^{13}$ cm$^{-2}$. The ratio of the two absorption lines yields a temperature estimate of 23 K.

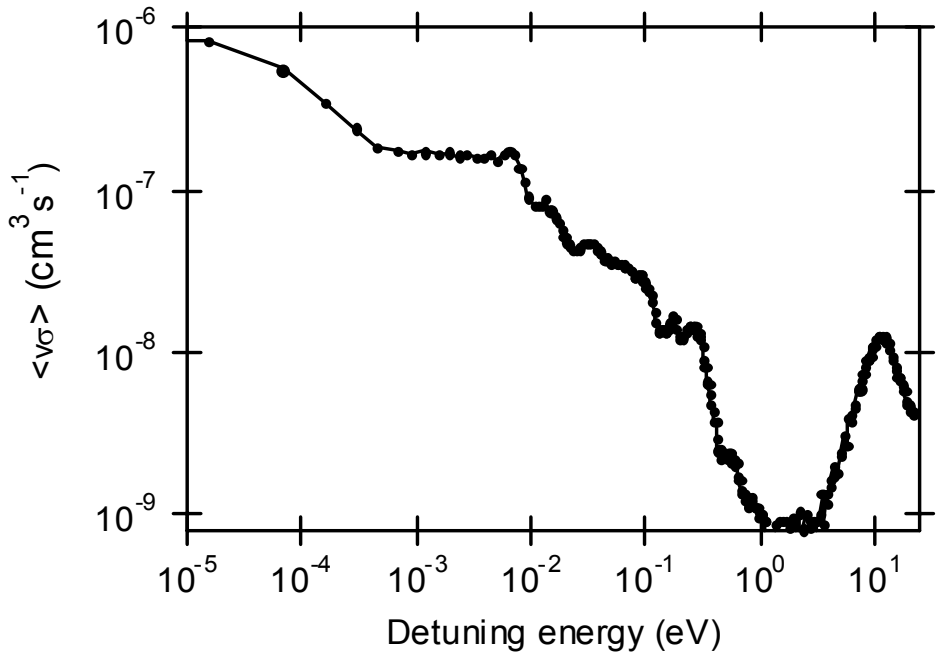

Figure 2: Measured dissociative recombination rate coefficient of rotationally cold $H_3^+$ as a function of detuning energy. For these measurements, the supersonic expansion ion source (described in Fig. 1a) was mounted onto the MINIS ion injection endstation of the CRYRING ion storage ring at the Manne Siegbahn Laboratory in Stockholm. The ions leaving the source were mass selected and accelerated to 300 keV/amu before being injected into the ring. After injection, the ions were accelerated to their final energy (12.1 MeV), which gave a circulating ion beam current of 0.31 µA. In one straight section ("interaction region") of the 52 m circumference storage ring, the ions passed collinearly through an approximately homogeneous electron beam of 85 cm length and 4 cm diameter ("electron cooler"). The electron beam current was 35.5 mA and the electron energy (2.2 keV) was chosen so that ion-electron velocity matching was achieved. For the first five seconds after the ion beam had reached its full energy, the electron beam acted to reduce the phase-space volume of the stored ion beam, leading to a narrow ion-velocity spread and a correspondingly small ion beam diameter (~1 mm). During this time, some stored ions were lost to DR with electrons and collisions with residual gas molecules, while the remaining ions relaxed to the ground vibrational state by spontaneous emission. This cooling phase was followed by a measurement of the total number of neutral reaction products (H + H + H and $H_2$ + H) using a surface barrier detector mounted tangentially to the ring following the interaction region. The electron energy was varied by changing the cathode voltage in the electron cooler, and the neutral fragment signal was recorded as a function of the longitudinal energy difference between electrons and ions ("detuning energy"). Note the resonance structure between 1 meV and 1 eV that was only hinted at in previous experiments using rotationally hot ions.

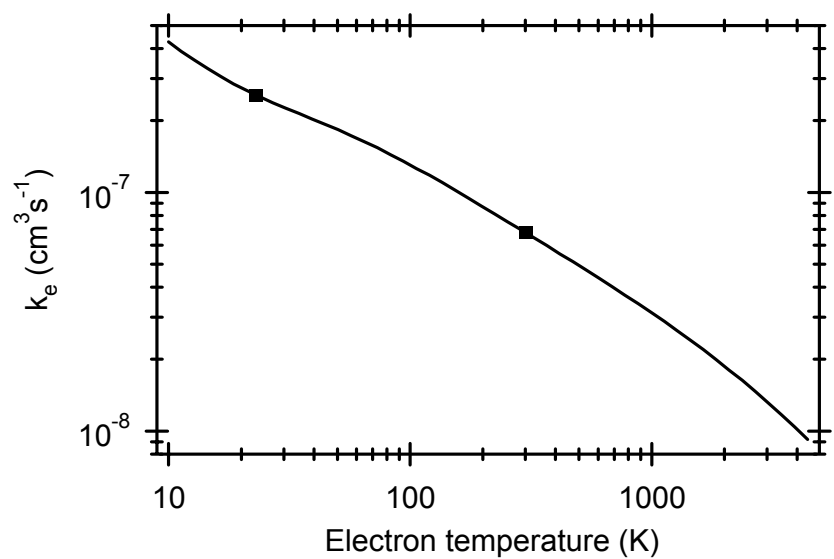

Figure 3: Calculated thermal rate coefficient $k_e$ for the dissociative recombination of rotationally cold $H_3^+$ ions as a function of electron temperature T, based on CRYRING measurements. The experimentally measured rate coefficient (<vσ>) shown in Fig. 2 was deconvolved using the known electron velocity distribution, yielding the dissociative recombination cross section σ as a function of electron energy. The cross section was then integrated over a Boltzmann distribution of energies (corresponding to a thermalized electron energy distribution) to yield the "thermal" rate coefficient at each value of the electron temperature. The values $k_e$(23 K)=2.6×$10^{-7}$ and $k_e$(300 K)=6.8×$10^{-8}$ $cm^3s^{-1}$ are indicated with squares. Note that the value of $k_e$(300 K) refers to an electron temperature of 300 K, but still pertains to rotationally cold $H_3^+$ ions. A more detailed manuscript with a thorough discussion of the experimental method and analysis is in preparation.